\documentclass[twocolumn,floatfix,superscriptaddress,a4paper,showpacs,showkeys,nofootinbib,reprint,prl]{revtex4-1}
\usepackage{epsfig}
\usepackage{latexsym}
\usepackage{xspace}
\usepackage[colorlinks=true,linktocpage=true,linkcolor=blue,citecolor=blue,allcolors=blue]{hyperref}
\usepackage[utf8]{inputenc}
\usepackage{indentfirst}
\usepackage{enumerate}
\usepackage{color}

\usepackage[caption=false,position=top]{subfig}

\usepackage{amsmath}
\usepackage{amssymb}
\usepackage[english]{babel}
\usepackage{url}

\newcommand{\eq}[1]{\begin{align} #1 \end{align}}
\newcommand{\mean}[1]{\langle #1 \rangle}
\newcommand{\cum}[1]{\kappa_{#1}}
\newcommand{\cumt}[1]{\tilde{\kappa}_{#1}}
\newcommand{\cumu}[1]{\kappa_{#1}[B_1]}
\newcommand{\dens}[1]{\rho_{#1}}

\begin{document}


\title{
Connecting fluctuation measurements in heavy-ion collisions\\ with the grand-canonical susceptibilities
}

\author{Volodymyr Vovchenko}
\affiliation{Nuclear Science Division, Lawrence Berkeley National Laboratory, 1 Cyclotron Road, Berkeley, CA 94720, USA}
\affiliation{
Institut f\"ur Theoretische Physik,
Goethe Universit\"at Frankfurt, Max-von-Laue-Str. 1, D-60438 Frankfurt am Main, Germany}
\affiliation{Frankfurt Institute for Advanced Studies, Giersch Science Center, Ruth-Moufang-Str. 1, D-60438 Frankfurt am Main, Germany}

\author{Oleh Savchuk}
\affiliation{Physics Department, Taras Shevchenko National University of Kyiv, 03022 Kyiv, Ukraine}

\author{Roman~V.~Poberezhnyuk}
\affiliation{Bogolyubov Institute for Theoretical Physics, 03680 Kyiv, Ukraine}
\affiliation{Frankfurt Institute for Advanced Studies, Giersch Science Center, Ruth-Moufang-Str. 1, D-60438 Frankfurt am Main, Germany}

\author{Mark I. Gorenstein}
\affiliation{Bogolyubov Institute for Theoretical Physics, 03680 Kyiv, Ukraine}
\affiliation{Frankfurt Institute for Advanced Studies, Giersch Science Center, Ruth-Moufang-Str. 1, D-60438 Frankfurt am Main, Germany}

\author{Volker Koch}
\affiliation{Nuclear Science Division, Lawrence Berkeley National Laboratory, 1 Cyclotron Road, Berkeley, CA 94720, USA}

\begin{abstract}
We derive the relation between cumulants of a conserved charge measured in a subvolume of a thermal system and the corresponding grand-canonical susceptibilities, taking into account exact global conservation of that charge. 
The derivation is presented for an arbitrary equation of state, with the assumption that the subvolume is sufficiently large to be close to the thermodynamic limit. 
Our framework -- the subensemble acceptance method~(SAM) -- quantifies the effect of global conservation laws and is an important step toward a 
direct comparison between cumulants of conserved charges measured in central heavy ion
collisions and theoretical calculations of grand-canonical susceptibilities, such as lattice QCD.
As an example, we apply our formalism to net-baryon fluctuations at vanishing baryon chemical
potentials as encountered in collisions at the LHC and RHIC. 
\end{abstract}


\keywords{fluctuations of conserved charges, conservation laws, heavy-ion collisions}

\maketitle


\paragraph*{Introduction.}

Studies of the QCD phase diagram are one of the focal points of  current experimental heavy-ion
collision programs~\cite{Bzdak:2019pkr}. 
Observables characterizing fluctuations of the QCD conserved charges -- baryon number, electric
charge, and strangeness -- have attracted a particular attention, as these are sensitive to the
finer details of the QCD equation of state and its phase structure in particular~\cite{Stephanov:1998dy,Stephanov:1999zu,Koch:2008ia}.
Consider, for simplicity, a case of a single conserved charge, say baryon number  $B$, for a system in equilibrium with volume $V$ at temperature $T$. The $n$th order scaled susceptibility $\chi_n^B$ is defined as a derivative of the pressure with respect to the chemical potential $\mu_B$,
\eq{\label{eq:gce}
\chi_n^B \equiv \frac{\partial^n(p/T^4)}{\partial (\mu_B/T)^n} = \frac{\cum{n}[B]}{V\,T^3},
}
and it determines the cumulants, $\cum{n}[B]$, of the distribution of the charge $B$ in the grand canonical ensemble~(GCE).
The susceptibilities, $\chi_n^B$, characterize the properties of the thermal system under
consideration, in particular they provide information about the possible phase changes, including
remnants of the chiral criticality at vanishing chemical potential~\cite{Friman:2011pf}.
Theoretically they are calculated either using first-principle lattice QCD simulations~\cite{Bazavov:2017dus,Borsanyi:2018grb}, or in various effective QCD approaches~\cite{Isserstedt:2019pgx,Fu:2019hdw}.
An important question is how to relate these quantities to experimental measurements~\cite{Bleicher:2000ek,Schuster:2009jv,Bzdak:2012an,Braun-Munzinger:2016yjz,Pruneau:2019baa}.
The total net charge $B$ does not fluctuate in the course of a heavy-ion collision, as opposed to the case of the GCE where the system can freely exchange the charge with an external heat bath.
However, experimental measurements typically have limited acceptance and  only cover a
fraction of the total momentum space, which we subsequently assume to be characterized by a finite
acceptance window in rapidity, $\Delta Y_{\rm acc}$. As discussed e.g. in \cite{Koch:2008ia}, for a
sufficiently small acceptance window $\Delta Y_{\rm acc}\ll \Delta Y_{\rm 4\pi} $ conditions corresponding to the GCE may be
imitated, i.e. effects of global charge conservation become negligible. However,  in order to
capture the relevant physics the acceptance window $\Delta Y_{\rm acc}$ must be much larger than
the correlation length $\Delta Y_{\rm cor}$. Consequently, 
$\Delta Y_{\rm cor} \ll \Delta Y_{\rm acc} \ll \Delta Y_{\rm 4\pi}$ is the minimum necessary condition for the applicability of the GCE to fluctuation measurements.

In practice the situation is more subtle.
Deviations of lattice QCD calculations of
the cumulants of conserved charges at $T \sim 160$~MeV from the ideal hadron resonance gas~(HRG) expectation do not exceed the magnitude of
charge conservation effects already for an acceptance as small as $\Delta Y_{\rm acc} / \Delta Y_{\rm 4\pi} \sim 0.1$~\cite{Bzdak:2012an}.
Therefore,
in order to capture the physics of e.g. chiral criticality the effect of charge conservation
needs to be understood very well, since simply reducing the acceptance window even further risks
eliminating all the non-trivial effects associated with relevant QCD dynamics~\cite{Ling:2015yau}.

In the present letter we generalize the relation (\ref{eq:gce}) between the GCE susceptibilities
$\chi_n^B$ and measured cumulants of conserved charge $\cum{n}[B]$ 
to make it valid for subsystems
that are comparable in size to the total system.
We will still assume that the size of the
subsystem is large enough to capture the relevant physics. Further assuming strong space-momentum
correlations, as is the case for LHC and top RHIC energies,
the formalism presented 
here connects the measured cumulants with those obtained in lattice QCD over a wide
range of acceptance windows.

\paragraph*{Formalism.}

\begin{figure}[t]
  \centering
  \includegraphics[width=.45\textwidth]{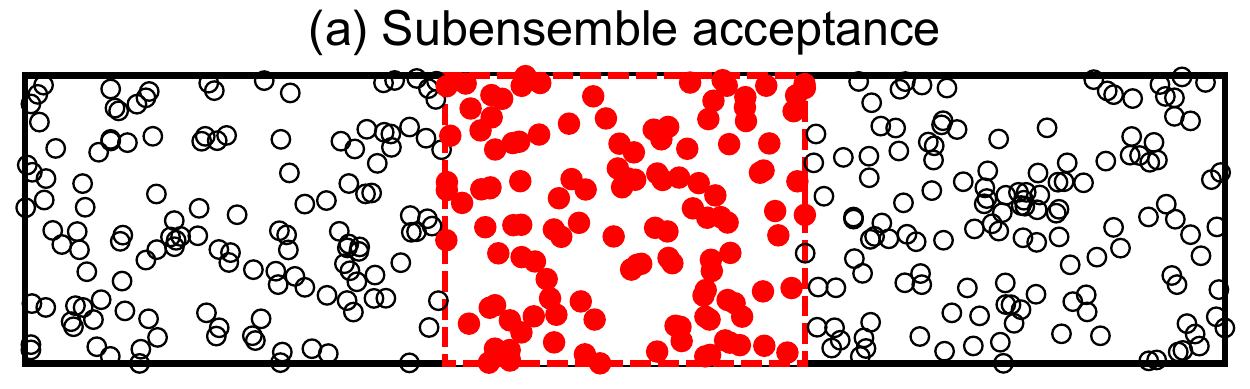}
  \vskip5pt
  \includegraphics[width=.45\textwidth]{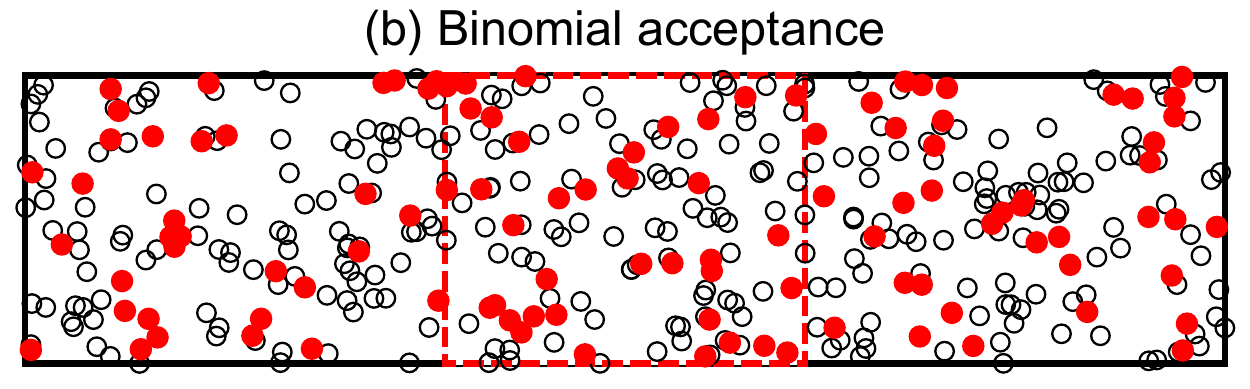}
  \caption{
A subsystem~(dashed red rectangle) within a thermal system~(solid black rectangle). The subsystem can exchange particles~(conserved charges), shown by the circles, with the rest of the system. 
The filled red circles in (a) depict the particles within the subsystem, as considered in the present subensemble acceptance method~(SAM).
In contrast, the filled red circles in (b) highlight the particles from a typical configuration resulting from the binomial filter.
  }
  \label{fig:subsyst}
\end{figure}

Consider a spatially uniform thermal system at a fixed temperature $T$, volume $V$, and total net
charge, say net baryon number, $B$, which is described by  statistical mechanics in the canonical ensemble and characterized by its canonical partition function $Z(T,V,B)$.
We pick a subsystem of a fixed volume $V_1 = \alpha \, V$ within the whole system, which can freely exchange the conserved charge $B$ with the rest of the system~(see Fig.~\ref{fig:subsyst}).
Our goal is to evaluate the cumulants $\cum{n}[B_1]$ of the distribution of charge $B_1$ within the subsystem~[the red points in Fig.~\ref{fig:subsyst}(a)].
Our considerations will extend the ideal HRG model results of Refs.~\cite{Bzdak:2012an,Braun-Munzinger:2016yjz} to arbitrary equations of state.
The subvolume cumulants in the ideal HRG model can be computed using a binomial filter, which corresponds to an independent acceptance of particles with a  probability $\alpha$ from the \emph{entire} volume $V$~[the red points in Fig.~\ref{fig:subsyst}(b)].
Given a finite correlation length, however, particles will be more strongly correlated with their neighboring particles than with those far away.
The binomial filter artificially suppresses these correlations and thus will not provide the correct results for the subvolume $V_1$.

Our arguments will be based purely on statistical mechanics. Assuming the subvolume $V_{1}$ as
well as the remaining volume $V_{2}=(1-\alpha)V$ to be large
compared to correlation length $\xi$, $V_{1} \gg \xi^{3}$ and $V_{2} \gg \xi^{3}$,  the canonical ensemble partition
function of the total system with total baryon number $B$ is given by
\eq{
Z(T,V,B)
  = \sum_{B_1} Z
  (T,\alpha V,B_{1}) Z
  (T,\beta V,B-B_{1}) \label{ZB}
}
Here $\beta \equiv 1 -\alpha$. 
The probability $P(B_1)$ to find $B_{1}$ baryons in the subsystem with volume $V_{1}$ is proportional to the product of the canonical partition functions of the two subsystems:
\eq{
P(B_1) &  \propto Z
(T,\alpha V,B_1) \, Z
(T,\beta V,B-B_1). \label{PB}
}
The procedure based on Eqs.~(\ref{ZB}) and (\ref{PB}) will be called the \emph{subensemble acceptance method}~(SAM). Note that the SAM reduces to the binomial acceptance sampling for the case of ideal HRG\footnote{Strictly
speaking, this is valid for the classical ideal HRG when quantum statistics effects can be
neglected. This is the case especially for baryons, where, due to their large mass,
corrections to baryon number cumulants arising from Fermi statistics are small  at the chemical
freeze-out, $T\simeq 155\,\rm{MeV}$ and $\mu_{B}\simeq 0$.}.

In the thermodynamic limit, i.e. for $V \to \infty$, the above results can be generalized, since in this case 
the canonical partition function can be 
expressed through the volume-independent free energy density $f$: $Z(T,V,B) = \exp\left[-\frac{V}{T} \, f(T,\dens{B}) \right]$ with $\dens{B} \equiv B/V$ being the conserved baryon density.
To evaluate $\cumu{n}$ we introduce the cumulant generating function $G_{B_1}(t)$:
\eq{
G_{B_1}(t) & \equiv \ln \mean{e^{t\,B_1}} =\ln \sum_{B_1} \exp(t B_1) P(B_1)\nonumber \\
& = \ln \left\{ \sum_{B_1} \, e^{t\,B_1} \, \exp\left[-\frac{\alpha V}{T} \, f(T,\dens{B_1}) \right] \right. \nonumber \\
& \quad \times \left. \exp\left[-\frac{\beta V}{T} \, f(T,\dens{B_2}) \right]  \right\} + \tilde{C}.
}
Here $\dens{B_2} = \frac{B-B_1}{V-V_1}$ is the charge density in the second subsystem and $\tilde{C}$ is
an irrelevant normalization constant. The cumulants, $\cumu{n}$, correspond to the Taylor coefficients of $G_{B_1}(t)$:
\eq{
\cum{n}[B_1] = \left. \frac{\partial^n G_{B_1}(t)}{\partial t^n} \right|_{t=0} \equiv \left.  \cumt{n}[B_1(t)] \right|_{t=0}.
}
Here we have introduced a shorthand,  $\cumt{n}[B_{1}(t)]$, for the n-th derivative of generating
function at arbitrary values of $t$, which we subsequently will refer to as $t$-dependent
cumulants. Clearly, all higher order cumulants are given as a $t$-derivative of the first order
$t$-dependent cumulant, $\cumt{1}[B_{1}(t)]$, which is given by 
\begin{align}
\cumt{1}[B_{1}(t)]= \frac{\partial G_{B_1}(t)}{\partial t}=  \frac{\sum_{B_1} \, B_1 \, \tilde{P}(B_1;t)}{\sum_{B_1} \, \tilde{P}(B_1;t)} = \mean{B_1(t)}
\end{align}
with the (un-normalized) $t$-dependent probability \eq{
\tilde{P}(B_1;t) = \exp\left\{t B_1 - V \, \frac{\alpha f(T,\dens{B_1}) + \beta f(T,\dens{B_2})}{T}\right\}.
}
In the thermodynamic limit, $V \to \infty$, $\tilde{P}$ 
has a sharp maximum at the mean value
of $B_{1}$, $\mean{B_1(t)}$ ~\cite{huang1987statistical}. 
The condition $\partial \tilde{P}(B_1;t) / \partial B_1 = 0$ determines the location of this maximum
resulting in an implicit relation that determines $\mean{B_1(t)}$:
\eq{\label{eq:Q1t}
t = \hat{\mu}_B[T,\dens{B_1}(t)] - \hat{\mu}_B[T,\dens{B_2}(t)]~.
}
Here $\hat{\mu}_B = \mu_B/T$, and $\dens{B_1}(t) = \mean{B_1(t)} /
(\alpha V)$, $\dens{B_2}(t) = [B - \mean{B_1(t)}] / [(1-\alpha)V]$. We also used the thermodynamic relation $[\partial f(T,\dens{B}) / \partial \dens{B}]_T = \mu_B(T,\dens{B})$.
It follows from Eq.~\eqref{eq:Q1t} that $\dens{B_1} = \dens{B_2} = B/V$ for $t = 0$, i.e. the net baryon number is uniformly distributed between the two subsystems, as it should be by construction.
Therefore, 
\eq{\label{eq:C1}
\cumu{1} = \alpha \, \cum{1}[B] = \alpha \, V T^3 \, \chi_1^B.
}
The second cumulant is given by the $t$-derivative of $\cumt{1}[B_1(t)]$, i.e. $\cumt{2}[B_1(t)] = \partial \cumt{1}[B_1(t)] / \partial t = \mean{B_1' (t)}$.
To determine $\mean{B_1' (t)}$ we differentiate Eq.~\eqref{eq:Q1t} with respect to $t$.
To evaluate the $t$-derivative of the r.h.s of.~\eqref{eq:Q1t} we apply the chain rule $\partial  \hat{\mu}_B / \partial t = [\partial  \hat{\mu}_B(T,\dens{B_{1,2}}) / \partial \dens{B_{1,2}}]_T \, \, [\partial  \dens{B_{1,2}}(t) / \partial t]$ and use a thermodynamic identity $[\partial  \hat{\mu}_B(T,\dens{B_{1,2}}) / \partial \dens{B_{1,2}}]_T = [T^3\,\chi_2^B(T,\dens{B_{1,2}})  ]^{-1}$~.
The solution for the resulting equation for $\mean{B_1' (t)} \equiv \cumt{2}[B_1(t)]$ is 
\eq{\label{eq:Q1deriv}
\cumt{2}[B_1(t)] = \frac{V \, T^3}{[\alpha \, \chi_2^B(T,\dens{B_1})]^{-1} + [\beta \, \chi_2^B(T,\dens{B_2})]^{-1}}~
}
which at $t = 0$ gives the 2nd order  cumulant
\eq{\label{eq:C2}
\cum{2}[B_1] = \alpha \, (1-\alpha) \, V \, T^3 \, \chi_2^B~.
}

In order to evaluate the higher-order cumulants $\cum{n}[B_1]$ for $n \geq 3$ we iteratively differentiate the $t$-dependent cumulants $\cumt{n}[B_1(t)]$ with respect to $t$, starting from $\cumt{2}[B_1(t)]$, and make use of the expression~\eqref{eq:Q1deriv} for $\mean{B_1' (t)}$.
The result for the cumulants up to the 6th order is the following:
\eq{
& \frac{\cum{3}[B_1]}{\alpha \, V \, T^3}  = \beta \, (1-2\alpha) \, \chi_3^B~, \\
& \frac{\cum{4}[B_1]}{\alpha \, V \, T^3}  = \beta \, \left[\chi_4^B~ - 3 \alpha \beta \frac{(\chi_3^B)^2 + \chi_2^B \, \chi_4^B}{\chi_2^B}\right], \\
& \frac{\cum{5}[B_1]}{\alpha \, V \, T^3}  = \beta \, (1-2\alpha) \left\{[1-2\beta\alpha]\chi_5^B~ - 10 \alpha \beta \frac{\chi_3^B  \chi_4^B}{\chi_2^B}\right\}, \\
\label{eq:C6}
& \frac{\cum{6}[B_1]}{\alpha \, V \, T^3}  = \beta \left[1-5\alpha \beta (1 - \alpha \beta ) \right]\chi_6^B + 5 \, \alpha \, \beta^2 \nonumber \\
& \quad \times \left\{9 \alpha \beta \frac{(\chi_3^B)^2 \, \chi_4^B}{(\chi_2^B)^2} - 3 \alpha \beta \frac{(\chi_3^B)^4}{(\chi_2^B)^3}  \right. \nonumber \\
& \quad \quad \left.  - 2 (1-2\alpha)^2 \frac{(\chi_4^B)^2 }{\chi_2^B} - 3[1 - 3\beta \alpha] \frac{\chi_3^B \, \chi_5^B}{\chi_2^B} \right\}.
}

In the limit $\alpha \to 0$ all susceptibilities, i.e. the cumulants scaled by $V_1 T^3 \equiv \alpha V
T^3$, reduce to the GCE susceptibilities, as expected, since in this limit effects of global
conservation become negligible.
Note, however, that the $\alpha \to 0$ limit discussed here assumes that the condition $V_1 \gg \xi^3$ still holds no matter how small the value of $\alpha$ is.
Such a scenario can be realized by holding the subsystem volume fixed to a sufficiently large value and increasing the total volume, i.e. $V_1 = \rm{const} \gg \xi^3$ and $V \to \infty$.

In heavy-ion collisions, on the other hand, a different scenario is realized.
The total volume is fixed while the volume of the subsystem is regulated by the measurement acceptance for example in longitudinal rapidity. 
This implies that the $\alpha \to 0$ limit corresponds to $V = \rm{const}$ and $V_1 \to 0$, meaning that our assumption of the subsystems being close to the thermodynamic limit breaks down, as the subsystem becomes much smaller than the correlation length, $\alpha V \ll \xi^{3}$. 
The cumulants then approach the Poisson limit \cite{Bzdak:2017ltv} rather than the GCE limit.
We return to the discussion of this point when we apply our method to net baryon fluctuations at the LHC and RHIC.
In the other limit, $\alpha \to 1$, all cumulants  of order $n \geq 2$ tend to zero, reflecting the dominance of the global conservation laws and the absence of conserved charge fluctuations in the full volume.

The derivations in the SAM assume that both volumes are much larger than the correlation length, i.e. $V_1,V_2 \gg \xi^3$.
While this condition is realized in many scenarios, one case where this may not hold is a vicinity of a critical point.
The correlation length diverges at the critical point, $\xi \to \infty$, thus the applicability of the SAM in its vicinity may be limited.
In the present work we will apply the formalism only at LHC and top RHIC energies where this issue is not relevant.

It is instructive to consider ratios of cumulants, in which the volume $V$ cancels. The explicit relations for the commonly used scaled variance, skewness, and kurtosis are:
\eq{\label{eq:w}
\frac{\cumu{2}}{\cumu{1}} & = (1-\alpha) \, \frac{\chi_2^B}{\chi_1^B}, \\
\label{eq:skew}
\frac{\cumu{3}}{\cumu{2}} & = (1-2\alpha) \, \frac{\chi_3^B}{\chi_2^B}, \\
\label{eq:kurt}
\frac{\cumu{4}}{\cumu{2}} & = (1-3\alpha \beta) \, \frac{\chi_4^B}{\chi_2^B} - 3 \alpha \beta \left( \frac{\chi_3^B}{\chi_2^B}\right)^2.
}

The modification of the scaled variance $\cumu{2}/\cumu{1}$ due to global conservation laws is a multiplication of the grand canonical scaled variance by a factor $(1-\alpha)$.
This is similar to the binomial filter effect studied in prior works~\cite{Bzdak:2012an,Braun-Munzinger:2016yjz,Savchuk:2019xfg}.
Same for the 
skewness $\cumu{3}/\cumu{2}$, where the corresponding grand canonical ratio is multiplied by $(1 - 2\alpha)$.
An interesting case is the kurtosis $\cumu{4}/\cumu{2}$: this ratio in the subvolume depends not
only on the GCE kurtosis $\chi_4^B/\chi_2^B$ but also on the GCE skewness $\chi_3^B/\chi_2^B$.
If $\alpha$ is known, Eqs.~\eqref{eq:w}-\eqref{eq:kurt} may be inverted to express the GCE cumulant ratios in terms of those of the subsystem.

\paragraph*{Net baryon fluctuations at LHC and top RHIC energies.}

We apply our formalism to study the effect of baryon number conservation in view of measurements of
net proton number distributions in heavy-ion collisions at the RHIC and LHC.
The ALICE collaboration has published measurements of the variance of net proton
distribution~\cite{Acharya:2019izy} and the analysis of higher orders up to $\kappa_4$ is in progress. 
In the future runs, sufficient statistics may be accumulated to extend the measurements up to the 6th order~\cite{Citron:2018lsq}.
The STAR collaboration has measured the cumulants of the net proton distribution up to $\kappa_4$~\cite{Adamczyk:2013dal,Adam:2020unf}, preliminary results for $\kappa_6$ are also available~\cite{Nonaka:2019fhk}.

It should be noted that experimental measurements in heavy-ion experiments are performed in  momentum space rather than in  coordinate space. 
However, the momenta and coordinates of particles at freeze-out are correlated due to the presence of a sizable collective flow, in particular the longitudinal flow.
The correlation is one-to-one in the case of a Bjorken
scenario, which can be expected to be approximately realized at the highest collision energies
achievable at the RHIC and LHC. In that case, the experimental momentum cuts in rapidity correspond to cuts in
coordinate space and our formalism is applicable,
provided that all transverse momenta are covered.\footnote{We note that thermal smearing by $\Delta Y_{\rm th} \sim (T/m)^{1/2}$ somewhat dilutes the space-momentum correlation~\cite{Ling:2015yau,Ohnishi:2016bdf}. In case of baryons, which are heavy, this effect is rather small, especially if a rapidity window of $\Delta Y_{\rm acc}\simeq 2$ is considered. We expect baryon smearing to slightly shift our results for the cumulant ratios towards that obtained using the binomial filter. 
} 
In the other extreme,  when no collective motion
is present,  cuts in the momentum space do not correlate with a definite subvolume in the coordinate space. 
In that case the binomial acceptance may be the appropriate procedure. 
We also note that our calculations apply to net baryon fluctuations rather than net proton ones. Experimentally, the former can be reconstructed from the latter following the binomial-like method developed in Refs.~\cite{Kitazawa:2011wh,Kitazawa:2012at}. 
This method requires the knowledge of various factorial moments, which cannot be obtained from statistical physics alone but can and should be measured in the experiment.

The typical chemical freeze-out temperatures, $T_{\rm ch} \sim
155-160$~MeV at the LHC~\cite{Andronic:2017pug,Becattini:2012xb,Petran:2013lja} and $T_{\rm ch} \sim 160-165$~MeV at the top RHIC energies~\cite{Adamczyk:2017iwn}, are close to the pseudo-critical
temperature of the QCD crossover transition determined by lattice QCD $T_{\rm pc} \simeq
155-160$~MeV~\cite{Bazavov:2018mes,Borsanyi:2020fev} at $\mu_B = 0$. Also, in the vicinity of
$T_{\rm pc}$   lattice calculations predict a change of sign of $\chi_6^B$, which is thought to be
related to the remnants of the chiral 
criticality~\cite{Friman:2011pf}, although alternative explanations do also
exist~\cite{Vovchenko:2016rkn,Vovchenko:2017gkg}. Therefore, it would be of great interest to verify the theory prediction of a negative $\chi_6^B$ experimentally.

\begin{figure*}[t]
  \centering
  \includegraphics[width=.49\textwidth]{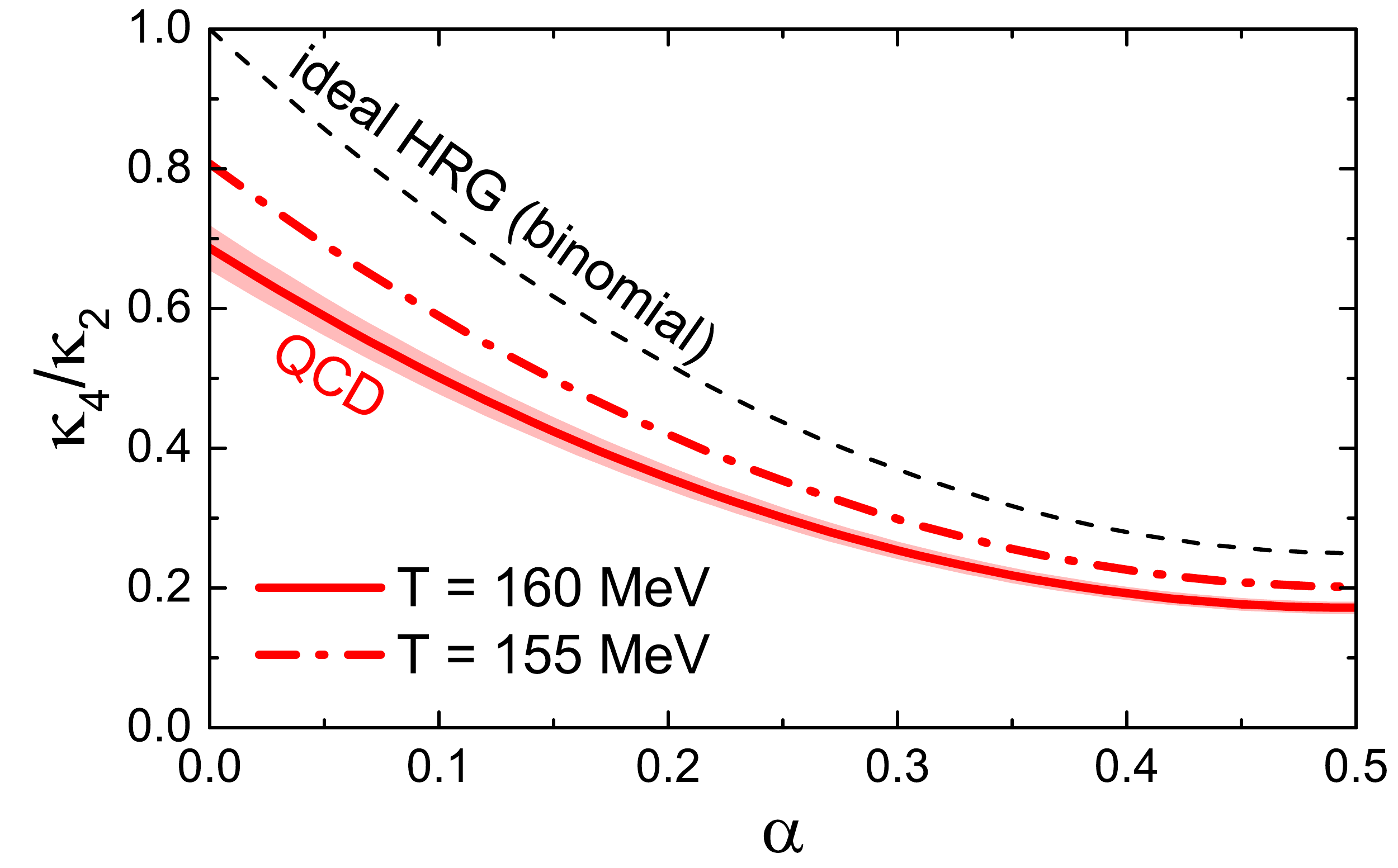}
  \includegraphics[width=.49\textwidth]{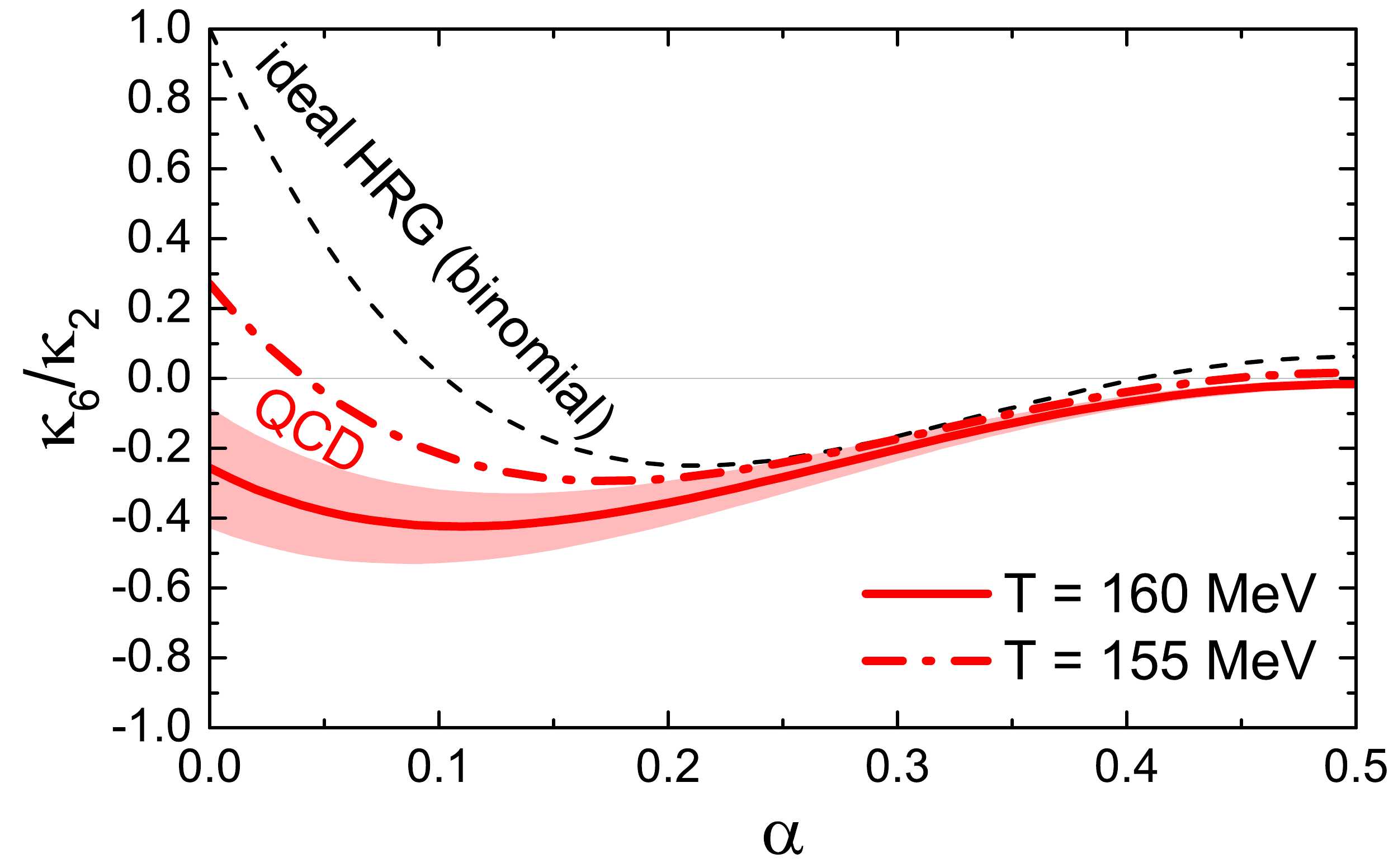}
  \caption{
   The dependence of $\cum{4}/\cum{2}$~(left) and $\cum{6}/\cum{2}$~(right) ratios calculated for net baryon fluctuations inside a subvolume on the fraction $\alpha$ of the total system volume covered for QCD matter at the LHC conditions at the chemical freeze-out at $T = 160$~MeV~(solid lines) and $T = 155$~MeV~(dash-dotted lines). 
   The bands correspond to the error propagation of the lattice data.
   The dashed lines depict the ideal hadron resonance gas results~(binomial filter).
   The results are symmetric with respect to a change $\alpha \to 1 - \alpha$ and thus shown up to $\alpha = 0.5$.
  }
  \label{fig:chiLHC}
\end{figure*}

As all odd order susceptibilities vanish at $\mu_B = 0$, the relations between the higher-order cumulants $\cumu{n}$ and susceptibilities $\chi_n^B$ simplify considerably.
For the kurtosis $\cumu{4}/\cumu{2}$ only the first term in Eq.~\eqref{eq:kurt} contributes.
The hyperkurtosis 
$\cumu{6}/\cumu{2}$ is obtained from Eqs.~\eqref{eq:C6} and \eqref{eq:C2},
\eq{\label{eq:kappa6}
\left(\frac{\cumu{6}}{\cumu{2}}\right)_{LHC} & = \left[1-5\alpha \beta (1 - \alpha \beta ) \right] \frac{\chi_6^B}{\chi_2^B} \nonumber \\
& \quad - 10 \alpha (1-2\alpha)^2 \beta \left( \frac{\chi_4^B}{\chi_2^B}\right)^2~.
}

We then study how the cumulant ratios $\cum{4}/\cum{2}$ and $\cum{6}/\cum{2}$ of net baryon distribution depend on
the value of the parameter $\alpha$ characterizing the subsystem where the fluctuations are
measured. We use the lattice data for $\chi_4^B/\chi_2^B$ and $\chi_6^B/\chi_2^B$ at $T=155$ and 160~MeV from Ref.~\cite{Borsanyi:2018grb} as input to the SAM.
The results for the $\alpha$-dependence of 
$\cum{4}/\cum{2}$ and $\cum{6}/\cum{2}$ ratios calculated from Eqs.~(\ref{eq:kurt}) and (\ref{eq:kappa6}) are shown in Fig.~\ref{fig:chiLHC}. 
The solid red lines correspond to $T = 160$~MeV and the bands depict the error propagation of the lattice data. The dash-dotted lines correspond to $T = 155$~MeV.
We also show the binomial acceptance results as dashed lines.
These results are only valid for the classical ideal HRG which
gives
$\chi_4^B/ \chi_2^B = 1$ and $\chi_6^B/\chi_2^B=1$. 

As both the kurtosis and hyperkurtosis are symmetric with respect to $\alpha \leftrightarrow (1 - \alpha)$~\cite{Bzdak:2017ltv}  we plot our results only up to $\alpha = 0.5$.
In the limit $\alpha \to 0$ both $\cum{4}/\cum{2}$ and $\cum{6}/\cum{2}$ approach their GCE values.
The computed values of $\cum{4}/\cum{2}$ and $\cum{6}/\cum{2}$ lie below the binomial acceptance baseline for all
values of $\alpha$, which 
reflects the suppression of the lattice values for $\chi_4^B/\chi_2^B$ and $\chi_6^B/\chi_2^B$ relative to the ideal HRG
baseline.
Interestingly, the difference between the ideal gas and QCD is the smallest for $\cum{6}/\cum{2}$ at
$\alpha = 0.5$, where the effects of baryon conservation are the strongest. Actually, in the entire region $0.2
< \alpha < 0.8$ the difference is so small that 
it may be difficult to distinguish  the true dynamics of QCD from that of an ideal HRG.
Measurements in this region of $\alpha$, on the other hand, 
may serve as a model-independent test of baryon number conservation effects.
For $\alpha < 0.2$, however, the measurable ratio $\cum{6}/\cum{2}$ become sensitive to the equation of state, i.e. to the actual value for $\chi_6^B/\chi_2^B$.
We find that a negative $\cum{6}/\cum{2}$ for $\alpha \lesssim 0.1$ is consistent with
$\chi_6^B/\chi_2^B$ which is either negative or close to zero. 
Such a measurement would constitute a potentially unambiguous experimental signature of the QCD chiral crossover transition.

If we apply these conditions to actual experiments such as ALICE and STAR, it translates into the
following: At the LHC (ALICE) with $\sqrt{s_{\rm NN}} = 5.02$~TeV, the beam rapidity is $y_{\rm beam} \simeq \ln(\sqrt{s_{\rm NN}}/m_N) \simeq 8.5$ while for the top RHIC energy ($\sqrt{s_{\rm NN}} = 200$~GeV) one has $y_{\rm beam} \simeq 5.4$.
Thus, $\alpha \lesssim 0.1$ would correspond to measurements within approximately two (one) units of rapidity for LHC~(RHIC). 

As discussed above, at $\alpha$ below a certain value,  $\alpha < \alpha_{\rm lim}$, our formalism breaks
down and the cumulants 
approach the Poisson limit instead.
The value of $\alpha_{\rm lim}$ can be estimated.
The physical volume used in lattice
calculations~\cite{Borsanyi:2011sw,Bazavov:2012jq,Bazavov:2017dus,Borsanyi:2018grb} of $\chi_{2n}^B$
at $T \simeq 160$~MeV is of order $V_{\rm lat} = (a \, N_{\sigma})^3 \sim 10^2$~fm$^3$ for $a = 0.1$~fm and $N_\sigma = 48$ lattices.
We can thus assume that volumes $V \geq V_{\rm lat}$ are sufficiently  large to capture the relevant physics.
The total volume in central collisions at the LHC can be estimated as $V_{\rm tot} = (dV / dY) \, 2 \, y_{\rm beam}$, which for $dV / dY \sim 5000$~fm$^3$~\cite{Andronic:2017pug} and $y_{\rm beam} \sim 8$ at the LHC yields $V_{\rm tot} \sim 80000$~fm$^3$.
Therefore, $\alpha_{\rm lim} = V_{\rm lat} / V_{\rm tot} \sim 10^{-3}$.
We note, however, that the shape of the freeze-out volume in heavy-ion collisions taken in a narrow space-time rapidity window is more resemblant of a disk rather than a squared box in lattice QCD.
This difference may introduce an error in our estimate of $\alpha_{\rm lim}$, meaning that the estimate likely lies on the optimistic side.
Nevertheless, even an order of magnitude error in this estimate implies that $\alpha_{\rm lim}$ does not exceed $10^{-2}$, and thus our method is applicable for virtually the entire linear scale shown in Fig.~\ref{fig:chiLHC}.
The same estimate for RHIC, where $dV/dy \simeq 1500$~fm$^3$~\cite{Adamczyk:2017iwn}, gives $\alpha_{\rm lim} \sim 7 \cdot 10^{-3}$.
Another important issue is the thermal smearing which dilutes the correlation between the space-time rapidity and the kinematic rapidity which is actually measured in experiment. 
The smearing induces a correlation length $\Delta Y_{\rm corr} \sim 1$ in kinematical rapidity~\cite{Ling:2015yau}, meaning that a rapidity acceptance of one unit or more may be required for this effect to be subleading. Measurements in $\Delta Y_{\rm acc}\simeq 2$ acceptance at the LHC will thus be less susceptible to the thermal smearing than $\Delta Y_{\rm acc}\simeq 1$ at RHIC. A detailed study of this correction is under way.

Our discussion corresponds to $p_T$-integrated measurements of higher-order net-proton fluctuations.
There are no conceptual problems preventing such measurements, however,
this has not yet been achieved in the presently available data collected at RHIC by STAR~\cite{Adamczyk:2013dal,Nonaka:2019fhk,Adam:2020unf} and at the LHC by ALICE~\cite{Acharya:2019izy}. In both experiments the acceptance covers only part of the whole $p_T$ range. In addition, ALICE data are restricted to second order cumulants, while the STAR data should be supplemented with measurements of factorial moments that, as mentioned earlier, are needed to recover baryon number cumulants.
These facts prevent us from analyzing the existing data within our formalism.
We do mention though, that the ALICE publication~\cite{Acharya:2019izy} has discussed baryon number conservation in the framework of the HRG model~(binomial acceptance), reporting indications for the relevance of the $(1-\alpha)$ factor~[Eq.~\eqref{eq:w}] due to baryon number conservation.

It should be noted that not only the baryon number is conserved in heavy-ion collisions, but the electric charge and strangeness as well. 
The SAM has been extended to the case of multiple conserved charges in Ref.~\cite{Vovchenko:2020gne}.
There it is shown that net-baryon fluctuations at $\mu_B = 0$ are affected by exact conservation of electric charge and strangeness only starting from the sixth order cumulant.
We verified this effect on $\cum{6}/\cum{2}$ within the ideal HRG model at $T = 160$~MeV and $\mu_B = 0$ and found deviations from Eq.~\eqref{eq:kappa6} to be negligibly small. 
Therefore, our results in Fig.~\ref{fig:chiLHC} are not expected to be 
affected significantly by the electric charge and strangeness conservation.

\paragraph{Summary.} 
We presented a novel procedure to connect measurements of cumulants of conserved charge fluctuations in a finite acceptance to the grand-canonical susceptibilities, taking into account effects due to exact charge conservation.
In contrast to prior works studying the ideal HRG model, our subensemble acceptance method works for an arbitrary
equation of state, under the assumption that the acceptance is sufficiently large to reach the
thermodynamic limit, and thus to capture all the relevant physics. 
The formalism is most suitable for central collisions of ultrarelativistic heavy-ion collisions at
the highest energies where we have a strong space-momentum correlations, and it enables  direct
comparisons between  experimental data on cumulants of conserved charges and theoretical
calculations of grand-canonical susceptibilities within effective QCD theories and lattice QCD
simulations. We consider our results  to be particularly helpful for the ongoing experimental effort to
study the QCD phase structure with fluctuation measurements. As a first application, we have
studied  the conditions under which a measurement of a net baryon hyperkurtosis $\cum{6}/\cum{2}$
can serve as an experimental signature of the QCD chiral crossover at $\mu_B = 0$, and we found a
rapidity window of $\Delta Y_{\rm acc}\simeq 2 (1)$ at LHC (RHIC) to be the sweet spot, where the QCD dynamics is not overshadowed by baryon number conservation effects.

Our framework opens a number of new avenues to explore. 
For instance, it can be interesting to test the limits of our approach in finite systems close to the critical
point, where the correlation length becomes comparable to the system size~\cite{Poberezhnyuk:2020ayn}.
Another extension is a simultaneous incorporation of multiple conserved charges~\cite{Vovchenko:2020gne}, enabling the analysis of the off-diagonal susceptibilities which is a relevant topic in light of the corresponding measurements that are being performed by the STAR collaboration at RHIC~\cite{Adam:2019xmk}.


\begin{acknowledgments}

\emph{Acknowledgments.} 
We thank A.~Bzdak for fruitful discussions.
V.V. was supported by the
Feodor Lynen program of the Alexander von Humboldt
foundation.
This work received support through the U.S. Department of Energy, 
Office of Science, Office of Nuclear Physics, under contract number 
DE-AC02-05CH11231231 and received support within the framework of the
Beam Energy Scan Theory (BEST) Topical Collaboration.

\end{acknowledgments}

\bibliography{semiGCE}


\end{document}